# Ideal Unconventional Weyl Point in a Chiral Photonic Metamaterial

Yihao Yang[1,2], Zhen Gao[1,#], Xiaolong Feng[3], Yue-Xin Huang[3], Peiheng Zhou[4], Shengyuan A. Yang[3,*], Yidong Chong[1,2,†], and Baile Zhang[1,2,‡]

[1]*Division of Physics and Applied Physics, School of Physical and Mathematical Sciences, Nanyang Technological University, 21 Nanyang Link, Singapore 637371, Singapore*

[2]*Centre for Disruptive Photonic Technologies, The Photonics Institute, Nanyang Technological University, 50 Nanyang Avenue, Singapore 639798, Singapore*

[3]*Research Laboratory for Quantum Materials, Singapore University of Technology and Design, Singapore 487372, Singapore*

[4]*National Engineering Research Center of Electromagnetic Radiation Control Materials, State Key Laboratory of Electronic Thin Film and Integrated Devices, University of Electronic Science and Technology of China, Chengdu 610054, China*

Unconventional Weyl points (WPs), carrying topological charge 2 or higher, possess interesting properties different from ordinary charge-1 WPs, including multiple Fermi arcs that stretch over a large portion of the Brillouin zone. Thus far, such WPs have been observed in chiral materials and acoustic metamaterials, but there has been no clean demonstration in photonics in which the unconventional photonic WPs are separated from trivial bands. We experimentally realize an ideal symmetry-protected photonic charge-2 WP in a three-dimensional topological chiral microwave metamaterial. We use field mapping to directly observe the projected bulk dispersion, as well as the two long surface arcs that form a noncontractible loop wrapping around the surface Brillouin zone. The surface states span a record-wide frequency window of around 22.7% relative bandwidth. We demonstrate that the surface states exhibit a novel topological self-collimation property and are robust against disorder. This work provides an ideal photonic platform for exploring fundamental physics and applications of unconventional WPs.

Weyl points (WPs) are point band degeneracies in three-dimensional (3D) momentum space where two bands linearly intersect [1-13]. A WP acts as a monopole of Berry flux in momentum space, carrying a topological charge (or Chern number) of ±1. As a direct physical consequence, open isofrequency arcs known as Fermi arcs appear on Weyl crystal surfaces, connecting the projections of oppositely charged WPs. These bandstructure features were first proposed for condensed matter systems [1,5,6,12], but classical-wave analogues were quickly developed [2,4]. In photonics, WPs have been experimentally observed in several different platforms, such as dielectric photonic crystals (PhCs) [3], metallic structures [7,11,14], magnetized semiconductors [13], and optical waveguide arrays [9,15]. Intriguing phenomena associated with photonic WPs include chiral zero modes [16], robust surface states [7,14], and novel electromagnetic scattering laws [10].

Recent studies have revealed the existence of novel classes of unconventional Weyl points (UWPs) with topological charge ≥2, which occur in chiral crystal structures without mirror, inversion, or other roto-inversion symmetry [17-29]. These UWPs are distinct from the conventional charge-1 WPs [3,9,11] as well as charge-0 Dirac points [30-32]. They include double WPs (twofold quadratically degenerate points with topological charge ±2) [7,18,22,28], spin-1 WPs (threefold linearly degenerate points with topological charge ±2) [24,25,27], triple WPs (twofold cubically degenerate points with topological charge ±3) [7,17], and charge-2 and charge-4 Dirac points (fourfold degenerate points with topological charge ±2 and ±4) [19,22,24,25,27,33]. UWPs have several unique features not possessed by ordinary WPs occurring in nonchiral crystals [2,3,11]. First, due to their high Chern number, multiple Fermi arcs emanate from them, in contrast with charge-1 WPs which are each connected to a single Fermi arc. Second, UWPs are generally pinned to high-symmetry momentum points enforced by the symmetries of the underlying chiral crystal, which usually causes their Fermi arcs to extend across the entire two dimensional (2D) surface Brillouin zone (BZ) [24,25]. These Fermi arcs can even form a noncontractible loop winding around the surface BZ [22,25,27]. By contrast, charge-1 WPs typically occur close to each other and are connected by short Fermi arcs [24,25]. Third, UWPs with opposite topological charges are typically separated in energy, allowing for a much wider energy window in which to host Fermi arcs [23].

So far, UWPs have been experimentally observed in condensed matter systems [23-25] and acoustic metamaterials [27,28]. Some intriguing phenomena have been associated with them, such as the quantized circular photogalvanic effect [34] and unconventional photocurrents [35]. In photonics, however, UWPs have proven challenging to realize. In a study of microwave-scale PhCs, Chen *et al.* predicted the existence of both charge-1 WPs and double WPs in the bandstructure, but their experiment was only able to access the former [7]. Recently, Vaidya *et al.* have developed a woodpile PhC hosting a double WP in the infrared regime [36]. However, this double WP co-exists with many trivial bands at the same frequency range, making it challenging to study in isolation; indeed, the topological surface-arc states were not accessed in that experiment.

Here, we report on the experimental observation of an ideal UWP in a 3D photonic topological chiral metamaterial operating in the microwave regime. The UWP has topological charge 2 and is protected by $C_3$ rotational symmetry and time-reversal symmetry. The topological surface states occur over a record-wide frequency window of relative bandwidth ≈ 22.7%, with no other bands occupying the frequency range. We use microwave measurements to map out the projected bandstructure, verifying the quadratic dispersion of the UWP as well as the existence of long surface arcs that form a noncontractible loop around the surface BZ. To our knowledge, such surface state features have never been observed in a photonic system. In addition, we show that the surface-arc states exhibit a phenomenon that we call "topological self-collimation", owing to their flat isofrequency contours. We also theoretically and experimentally verify the robustness of topological surface states to disorder.

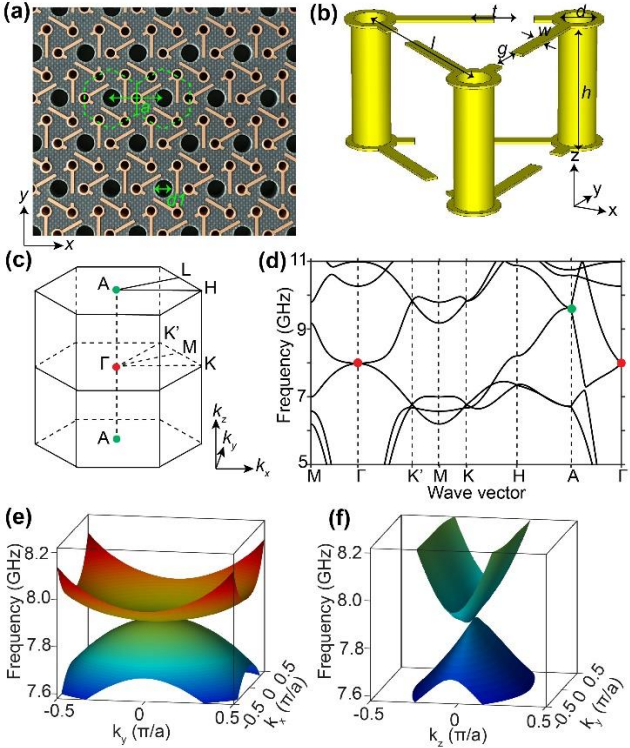

FIG. 1. Ideal UWP in a 3D topological chiral metamaterial. (a)-(b) Fabricated sample and unit cell of the 3D topological chiral metamaterial. (c) 3D first Brillouin zone (BZ). (d) Bandstructure of the chiral metamaterial. The red dot denotes the UWP at $\Gamma$, which has topological charge +2. The green dot represents the UWP at A, which has topological charge -2. (e)-(f) Perspective view of the 2D bandstructures in the vicinity of the UWP at $\Gamma$ in the $k_xk_y$ plane (e) and $k_yk_z$ plane (f).

The 3D topological chiral metamaterial is made from printed circuit boards (PCBs), as shown in Fig. 1(a). As shown in Fig. 1(b), it has a 3D hexagonal unit cell with in-plane lattice constant $a$=8.7 mm, consisting of three connected double-slit split ring resonators (SRRs) [37] made of copper; the structural parameters are $l$=7.8 mm, $g$=1.5 mm, $t$=1.5 mm, $w$=0.5 mm, $h$=5 mm, $d$=1.5 mm, and the thickness of the copper film is 0.035 mm. The slits of SRR are shifted by $t$ away from the centers of the triangle sides, which breaks all inversion and mirror symmetries. The resulting structure has chiral point group symmetry No. 150 ($\Gamma_h D_3^2$). The SRRs are embedded in a dielectric background with relative permittivity 2.7, and each PCB layer is paired with a blank layer of thickness 2 mm, so that the periodicity in $z$ direction is $p_z$ =7 mm. We introduce a lattice of air holes of diameter $d_1$=3.5 mm, passing through the triangle centers, for the insertion of microwave probes.

The BZ and bandstructure of the designed 3D topological chiral metamaterial is shown in Figs. 1(c-f). At the time-reversal-invariant momenta $\Gamma$ and A, the third and fourth bands intersect quadratically in the $k_x$ and $k_y$ directions and linearly in the $k_z$ direction [17,18]. These band-crossing points are UWPs with topological charge $\pm 2$ [7,17,18,28]. For the UWP at $\Gamma$, there are no additional trivial bands coexisting at the same frequency, so this is what previous authors have dubbed an "ideal WP" [11,21,24,38]. We numerically verified that the topological charges of the UWPs at $\Gamma$ and A are +2 and -2 respectively, using Wilson loop calculations on spherical surfaces enclosing them in momentum space [26,27,30] (see Supplemental Material). The UWPs are stabilized by the $C_3$ rotational symmetry and time-reversal symmetry [7], as revealed by an effective Hamiltonian model (see Supplemental Material). To further confirm this analysis, we have formulated a tight-binding model with the same space group, which exhibits similar UWPs in its bandstructure (see Supplemental Material).

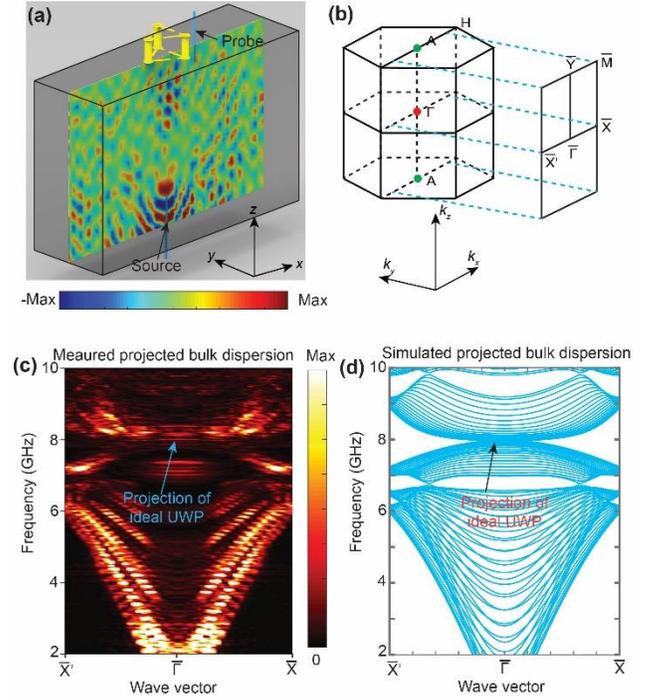

FIG. 2. Observation of the UWP. (a) Experimental setup. Field patterns in the middle $xz$ plane of the sample are measured data. (b) Schematic of the projected 2D surface BZ. (c) Measured projected bulk bandstructure along the high-symmetry line $\overline{X'X}$. The colormap measures the energy density. (d) Numerically calculated projected bulk bandstructure along the high-symmetry line $\overline{X'X}$.

We performed several experiments to characterize the topological chiral metamaterial. First, as shown in Fig. 2(a), we measure the projected bulk dispersion by inserting an electric dipole antenna (which acts as a point source) from the bottom of the sample. A second electric dipole antenna, acting as a probe, is inserted from the top, and we then map the complex field patterns within the central cutting plane of the sample.

The measured region is 435 mm by 350 mm with resolution 8.7 mm by 7.0 mm.

Applying 2D spatial Fourier transform to the measured complex field patterns at different frequencies, we obtain the bulk bandstructure projected onto the $k_x k_z$ plane, as indicated in Fig. 2(b). The results, shown in Fig. 2(c), reveal that the projected bulk bandstructure along the high-symmetry line $\overline{X'X}$ hosts a quadratic band-crossing at around 8.0 GHz. The measurement results are in good agreement with the computed projected bulk bandstructure shown in Fig. 2(d).

Next, we measure the topological surface states on an interface between the sample and air, as shown in Fig. 3(a). A point source is placed at the edge of the bottom surface (010), with the aim of exciting surface waves. The measured field pattern at 9.0 GHz is shown as an insert in Fig. 3(a). We then apply a Fourier transform to obtain the surface dispersion. The insert in Fig. 3(b) shows the momentum space distribution corresponding to the field pattern shown in Fig. 3(a). The two open arcs are the photonic Fermi arcs connecting the projections of the oppositely-charged UWPs at $\overline{\Gamma}$ and $\overline{Y}$.

Interestingly, the field pattern in Fig. 3(a) exhibits self-collimation: the waves emitted by the point source propagate along the surface in a tight beam that does not broaden with distance. This can be ascribed to the flat isofrequency contours of the photonic Fermi arcs. Similar self-collimation phenomena have previously been observed in topologically trivial PhCs [39-41], but in this case it arises from surface states that are topologically protected, meaning that they are robust against disorder that does not break translation symmetry along $z$ (see below). Self-collimation based on topologically protected surface waves may have future applications in photonics as a way of achieving reflection-free and diffraction-free light transport.

The measured surface dispersion along the high-symmetry line $\overline{\Gamma Y}$ is shown in Fig. 3(c). We see a surface sheet connecting two UWPs at separate frequencies. The sheet occupies an exceptionally large frequency window stretching from 7.8 GHz to 9.8 GHz (relative bandwidth of 22.7%), exceeding the bandwidths of all previously demonstrated Weyl photonic media [3,7-9,11,13]. The measured surface dispersion agrees well with numerical predictions, which are shown in Fig. 3(d). We plot the measured 2D surface isofrequency contours as a function of frequency in Fig. 3(e), which shows excellent agreement with the numerical predictions shown in Fig. 3(f). The number of surface sheets is consistent with the topological charges of the UWPs. The isofrequency contours of the two surface arcs between $\overline{\Gamma}$ and $\overline{Y}$ form a single noncontractible loop that wraps around the 2D surface BZ torus, a striking feature that has been observed in other forms of Weyl media but never before in photonics [22,25,27].

Next, we study the robustness of the surface states to a specific form of disorder consisting of randomly varying the gap sizes (denoted by $g$) on each column of SRRs, as shown in Figs. 4(a-b). The values of $g$ are drawn uniformly from 0.25 mm to 2.75 mm, with the SRRs in different layers of each column (i.e., at different $z$) having the same $g$. In the context of a single unit cell with periodic boundary conditions, varying $g$ shifts the frequencies of the UWPs at both $\Gamma$ and A, as shown in Fig. 4(c) (in the tight-binding model, this is equivalent to altering the on-site potential). As the WP frequencies are shifted by a frequency range of ~2.5 GHz, larger than the topological frequency window (~2 GHz), the disorder is considered strong.

We then measure the field distributions on the disordered sample, using the same experimental setup as before. As shown in Fig. 4(d), a point-like 9.3 GHz source is placed at the bottom of the disordered sample's (010) surface. A self-collimated surface beam is emitted from the source, despite the presence of strong disorder, appearing very similar to the disorder-free case shown in Fig. 3(a). Strikingly, the beam traverses a 90° sample edge experiencing negligible reflection. To further validate the robustness of this surface transport, we apply a 2D spatial Fourier transform to the complex electric field pattern measured on the (010) surface of the sample. The result, shown in Fig. 4(e), has an arc-like feature highly similar to the positive-$k$ surface arc in the disorder-free sample [Fig. 3(e)]. Notably, there is no backward-sloping arc that would correspond to a secondary beam reflected at the sample edge. From this, we conclude that the surface states are highly robust against this form of strong disorder.

The above results can be further understood as follows. Due to the discrete translational symmetry in $z$, the 3D system can be regarded as a collection of 2D subsystems of different $k_z$. These subsystems are equivalent to 2D photonic Chern insulators with the same in-plane disorder but different values of the $k_z$ parameter, which represents the strength of T breaking (analogous to a Haldane flux). For sufficiently large $k_z$, the 2D Chern insulators host bandgaps with chiral topological edge states that survive even in the regime of strong disorder [42]. The experimental results shown in Fig. 4(d,e) are consistent with this interpretation.

We have thus experimentally observed in a topological 3D chiral metamaterial an ideal symmetry-enforced photonic UWP, which is accompanied by two long Surface arcs spanning an extraordinarily wide frequency window. This platform provides a good opportunity to investigate applications and novel physical phenomena associated with UWPs. For example, the demonstrated topological self-collimation of the surface waves can be useful for implementing a reflection- and diffraction-free photonic medium. We also envision that exotic chiral zero modes and novel electromagnetic scattering laws can arise in such photonic Weyl media. Finally, it would be interesting to investigate the effects of optical activity and circular dichroism [43-45] in these 3D topological chiral metamaterials.

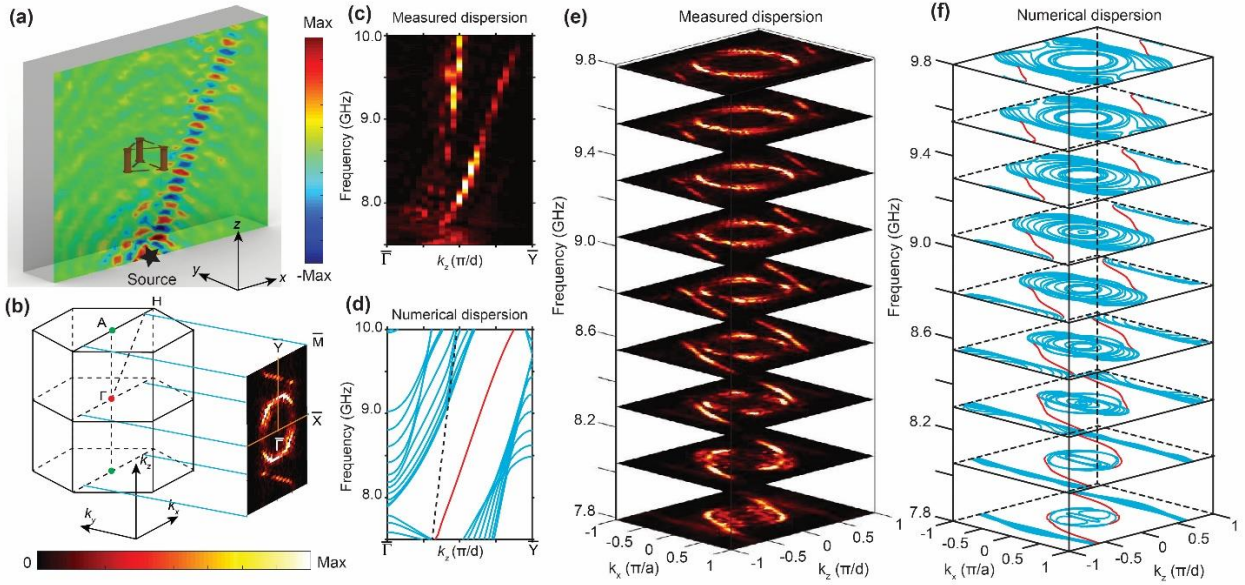

FIG. 3. Observed surface states of the ideal UWP, whose isofrequency contours form noncontractible loops wrapping around the BZ. (a) Experimental setup. The electric field distribution is mapped on the sample surface, with a source placed at the edge of the bottom surface (010). The inset image shows measurement data at 9.0 GHz, revealing a self-collimated beam of surface waves propagating from the source. (b) Schematic of the BZ projection onto the 2D measurement plane. The UWPs at Γ and A are projected onto $\bar{\Gamma}$ and $\bar{Y}$ respectively. The right inset shows measurement results obtained by Fourier transforming the complex field pattern measured in (a). (c)-(d) Measured and numerically calculated surface state dispersion along the high-symmetry line $\overline{\Gamma Y}$. The blue, red, and dotted black lines in (d) represent bulk dispersion, surface dispersion, and light line, respectively. (e)-(f) Measured and numerically calculated 2D isofrequency contours of the surface states as a function of frequency. The blue and red lines in (f) denote bulk and surface isofrequency contours, respectively. The circles in (a) near the BZ center are the air isofrequency contours. The color bar below (b) indicates the energy intensity.

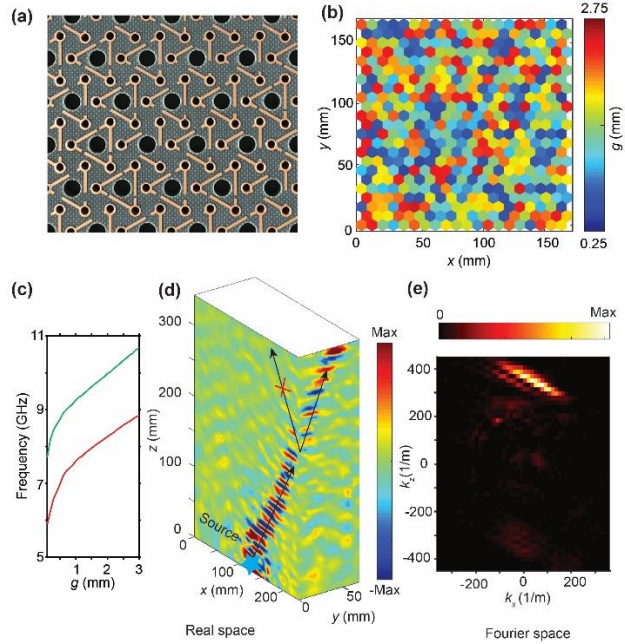

FIG. 4. Robust topological surfaces state. (a) The disordered chiral metamaterial. (b) Spatial distribution of $g$ in the $xy$ plane for the disordered sample. For each in-plane lattice site, $g$ is drawn randomly from 0.25 mm to 2.75 mm, and is independent of $z$. (c) Frequencies of the UWPs versus the gap shift parameter $g$. The red and green lines represent the UWPs at Γ and A, respectively. (d) Measured field distributions on the sample surfaces (010) and (100) at 9.3 GHz. The blue star denotes the location of a point-like source. (e) Energy maps obtained by 2D spatial Fourier transforming the measured field distributions on the sample surface (010).

This work was supported by the Singapore Ministry of Education under Grants No. MOE2016-T3-1-006, MOE2018-T2-1-022 (S) and MOE2019-T2-1-001.

#gaoz0008@e.ntu.edu.sg
*shengyuan_yang@sutd.edu.sg
†yidong@ntu.edu.sg
‡blzhang@ntu.edu.sg